\documentstyle[12pt]{article}
\newcommand{\Z}{{\sf Z \!\!\! Z}}
\newcommand{\R}{{\sf I \!\! R}}

\input{psfig}
\makeatletter
\@addtoreset{equation}{section}
\makeatother
 
\title{Complete Wetting in Supersymmetric QCD or \\ Why QCD Strings Can End on
Domain Walls 
\footnote{This work is supported in part by funds provided by the U.S.
Department of Energy (D.O.E.) under cooperative research agreement
DE-FC02-94ER40818.}}
\author{A. Campos, K. Holland and U.-J. Wiese \\ \\
Center for Theoretical Physics, \\
Laboratory for Nuclear Science, and Department of Physics \\
Massachusetts Institute of Technology (MIT) \\
Cambridge, Massachusetts 02139, U.S.A. \\ \\
MIT Preprint, CTP 2741 \\ \\}
 
\begin{document}
\maketitle
\begin{abstract} \normalsize
 
As argued by Witten on the basis of M-theory, QCD strings can end on domain 
walls. We present an explanation of this effect in the framework of effective 
field theories for the Polyakov loop and the gluino condensate in $N=1$ 
supersymmetric QCD. The domain walls separating confined phases with different 
values of the gluino condensate are completely wet by a layer of deconfined 
phase at the high-temperature phase transition. As a consequence, even at low 
temperatures, the Polyakov loop has a non-vanishing expectation value on the 
domain wall. Thus, close to the wall, the free energy of a static quark is 
finite and the string emanating from it can end on the wall.

\end{abstract}
 
\maketitle
 
\newpage

Recently, Witten has argued that QCD strings can end on domain walls 
\cite{Wit97}. This effect follows from a calculation in M-theory, where 
domain walls are represented by D-branes on which strings can end. Witten's 
argument applies to a theory in the universality class of $N=1$ supersymmetric 
QCD. In this theory, a $\Z(3)_\chi$ chiral symmetry --- an unbroken remnant of 
the anomalous $U(1)_R$ symmetry --- is spontaneously broken at low temperatures
by a non-zero value of the gluino condensate $\chi$. As a consequence, there 
are three distinct confined phases, characterized by three different values of 
$\chi$, which are related by $\Z(3)_\chi$ transformations. Regions of space 
filled with different confined phases are separated by domain walls. Passing 
from one region to another through a domain wall, the QCD vacuum angle $\theta$
changes by $2 \pi$. This behavior is characteristic of axionic domain walls.
Witten's string theory arguments imply that a QCD string emanating from a 
static quark can end on such a wall. The properties of domain walls in 
supersymmetric theories have been investigated in great detail by Shifman and
collaborators \cite{Kov97,Kog98} and topological defects ending on other 
defects have been studied in \cite{Car98}.

A qualitative explanation of why strings can end on axionic domain walls has 
been given by Rey \cite{Wit97} in the framework of field theory. His 
explanation is based on 't Hooft's picture of oblique confinement \cite{tHo81},
which at $\theta=0$ is due to monopole condensation. At $\theta \neq 0$ the 
monopoles turn into color-electrically charged dyons \cite{Wit79} and 
confinement is due to condensation of dyons. Hence, a confined-confined domain 
wall separates regions of monopole and dyon condensates. When a dyon passes 
through the domain wall, it turns into a monopole and leaves its color charge 
at the wall. If the walls can carry a color charge, a QCD string can end there.
A more quantitative description of this effect in an axion toy model has been 
given in \cite{Kog98}. While Rey's picture applies to the axion toy model, we
think that it does not provide a correct field theoretic explanation of 
Witten's M-theory calculation. We find that, in $N=1$ supersymmetric QCD, 
strings can end on the walls not because the walls are charged, but because 
they can transport flux to infinity. A similar scenario involving a non-Abelian
Coulomb phase was discussed qualitatively in \cite{Kog98}.

In this paper, we provide the first quantitative field theoretic explanation 
for why strings can end on walls for models in the universality class of $N=1$ 
supersymmetric QCD, i.e. for the same class of models to which Witten's 
M-theory calculation applies. Often, topological defects, which result from 
spontaneous symmetry breaking, have the phase of unbroken symmetry at their 
cores. For example, magnetic monopoles have symmetric vacuum at their centers. 
Similarly, at the core of axionic cosmic strings the $U(1)_{PQ}$ Peccei-Quinn 
symmetry is restored. Witten has constructed a model in which $U(1)_{PQ}$ 
symmetry restoration leads to breaking of the $U(1)_{em}$ gauge group of 
electromagnetism \cite{Wit85}. Hence, axionic strings in this model become 
superconducting. We proceed similarly for domain walls by constructing an 
effective field theory, in which $\Z(3)_\chi$ restoration implies breaking of
the $\Z(3)_c$ center symmetry of the $SU(3)$ gauge group. Then the 
high-temperature deconfined phase appears at the center of the domain wall, 
i.e. the Polyakov loop --- as the order parameter for $\Z(3)_c$ breaking ---
has a non-zero expectation value there. This means that a static quark has a 
finite free energy close to the wall and its string can end there.

The effect of a new phase appearing at a domain wall separating two bulk phases
is well-known in condensed matter physics as complete wetting. For example, the
interface between the human eye and the surrounding air is completely wet by a 
film of tears. Consequently, the eye-air solid-gas interface splits into a 
solid-liquid and a liquid-gas interface. The alternative is incomplete wetting.
Then droplets of liquid form at the solid-gas interface, but no complete 
wetting film appears. Complete wetting is known to appear at the deconfinement 
phase transition of the $N=0$ non-supersymmetric $SU(3)$ Yang-Mills theory 
\cite{Fre89,Tra92}. In that case, the $\Z(3)_c$ center symmetry is 
spontaneously broken at high temperatures, giving rise to three distinct 
deconfined phases, which are distinguished by different values of the Polyakov 
loop. When a deconfined-deconfined domain wall is cooled down to the phase 
transition, the low-temperature confined phase appears as a complete wetting 
layer that splits the deconfined-deconfined domain wall into a pair of 
confined-deconfined interfaces. In $N=1$ supersymmetric QCD, there are three 
distinct confined phases. In this paper, we show that complete wetting also 
arises when a confined-confined domain wall is heated up to the phase 
transition. In this case, one of the three deconfined phases forms a complete
wetting layer such that the Polyakov loop is non-zero at the domain wall. It
continues to be non-zero even below the phase transition temperature. As a
consequence, the free energy of a static quark located at the wall is finite. 
When the quark is displaced from the wall, its free energy increases linearly 
with the distance. Thus, the quark is confined to the wall and the string 
emanating from it ends there.

Complete wetting is a universal phenomenon of interfaces at first order phase
transitions. In the case of $N=0$ non-supersymmetric $SU(3)$ Yang-Mills theory
the universal aspects of the interface dynamics are captured by a 3-d effective
action \cite{Tra92},
\begin{equation}
S[\Phi] = \int d^3 x \ [\frac{1}{2} \vec \nabla \Phi^* \cdot \vec \nabla \Phi +
V(\Phi)],
\end{equation}
for the Polyakov loop $\Phi(\vec x)$, which is a gauge invariant complex scalar
field. Its expectation value $\langle \Phi \rangle \propto \exp(- F/T)$, where 
$T$ is the temperature, measures the free energy $F$ of a static quark. In the 
confined phase, $F$ diverges and $\langle \Phi \rangle$ vanishes, while in the 
deconfined phase, $F$ is finite and $\langle \Phi \rangle$ is non-zero. Under 
topologically non-trivial gauge transformations, which are periodic in 
Euclidean time up to a center element $z \in \Z(3)_c = \{e^{2 \pi i n/3}, 
n = 1,2,3\}$, the Polyakov loop changes into $\Phi' = \Phi z$. Hence, the 
$\Z(3)_c$ symmetry is spontaneously broken in the deconfined phase. Under 
charge conjugation, the Polyakov loop is replaced by its complex conjugate. The
effective potential $V(\Phi)$ is restricted by $\Z(3)_c$ and charge conjugation
symmetry, i.e.
\begin{equation}
V(\Phi z) = V(\Phi), \ V(\Phi^*) = V(\Phi).
\end{equation}
The most general quartic potential consistent with these symmetries takes the
form
\begin{equation}
V(\Phi) = a |\Phi|^2 + b \Phi_1(\Phi_1^2 - 3 \Phi_2^2) + c |\Phi|^4,
\end{equation}
where $\Phi = \Phi_1 + i \Phi_2$. One can restrict oneself to quartic 
potentials because they are sufficient to explore the universal features of the
interface dynamics. At the deconfinement phase transition temperature 
(corresponding to $b^2 = 4ac$), the above potential has four degenerate minima,
$\Phi^{(1)} = \Phi_0 \in \R$, $\Phi^{(2)} = (-1/2 + i\sqrt{3}/2) \Phi_0$ and 
$\Phi^{(3)} = (-1/2 - i\sqrt{3}/2) \Phi_0$ representing the three deconfined 
phases and $\Phi^{(4)} = 0$ representing the confined phase. In 
ref.\cite{Tra92}, it was shown that a deconfined-deconfined domain wall is 
completely wet by the confined phase and the corresponding critical exponents 
were determined analytically.

In $N=1$ supersymmetric QCD the $\Z(3)_\chi$ chiral symmetry is spontaneously
broken in the confined phase. The corresponding order parameter is the complex
valued gluino condensate $\chi = \chi_1 + i \chi_2$. Under chiral 
transformations $z \in \Z(3)_\chi$, the gluino condensate transforms into 
$\chi' = \chi z$ and under charge conjugation it also gets replaced by its 
complex conjugate. At high temperatures, one expects chiral symmetry to be 
restored and --- as in the non-supersymmetric theory --- the $\Z(3)_c$ center 
symmetry to be spontaneously broken due to deconfinement. Consequently, the 
effective action describing the interface dynamics now depends on both order 
parameters $\Phi$ and $\chi$, such that
\begin{equation}
S[\Phi,\chi] = \int d^3 x \ [\frac{1}{2} \vec \nabla \Phi^* \cdot \vec \nabla 
\Phi + \frac{1}{2} \vec \nabla \chi^* \cdot \vec \nabla \chi + V(\Phi,\chi)].
\end{equation}
The most general quartic potential consistent with $\Z(3)_c$, $\Z(3)_\chi$ and
charge conjugation now takes the form
\begin{equation}
V(\Phi,\chi) = a |\Phi|^2 + b \Phi_1(\Phi_1^2 - 3 \Phi_2^2) + c |\Phi|^4 +
d |\chi|^2 + e \chi_1(\chi_1^2 - 3 \chi_2^2) + f |\chi|^4 + 
g |\Phi|^2 |\chi|^2.
\end{equation}
First, we assume that deconfinement and chiral symmetry restoration occur at
the same temperature and that the phase transition is first order. Then, three
chirally broken confined phases coexist with three distinct chirally symmetric 
deconfined phases. The three deconfined phases have $\Phi^{(1)} = \Phi_0$, 
$\Phi^{(2)} = (-1/2 + i\sqrt{3}/2) \Phi_0$ and $\Phi^{(3)} = 
(-1/2 - i\sqrt{3}/2) \Phi_0$ and $\chi^{(1)} = \chi^{(2)} = \chi^{(3)} = 0$, 
while the three confined phases are characterized by $\Phi^{(4)} = \Phi^{(5)} =
\Phi^{(6)} = 0$ and $\chi^{(4)} = \chi_0 \in \R$, $\chi^{(5)} = 
(-1/2 + i\sqrt{3}/2) \chi_0$ and $\chi^{(6)} = (-1/2 - i\sqrt{3}/2) \chi_0$. 
The phase transition temperature corresponds to a choice of parameters 
$a,b,...,g$ such that all six phases $\Phi^{(n)}$, $\chi^{(n)}$ represent 
degenerate absolute minima of $V(\Phi,\chi)$.

We now look for solutions of the classical equations of motion, representing
planar domain walls, i.e. $\Phi(\vec x) = \Phi(z)$, $\chi(\vec x) = \chi(z)$, 
where $z$ is the coordinate perpendicular to the wall. The equations of motion 
then take the form
\begin{equation}
\frac{d^2 \Phi_i}{dz^2} = \frac{\partial V}{\partial \Phi_i}, \
\frac{d^2 \chi_i}{dz^2} = \frac{\partial V}{\partial \chi_i}.
\end{equation}
\begin{figure}
\psfig{figure=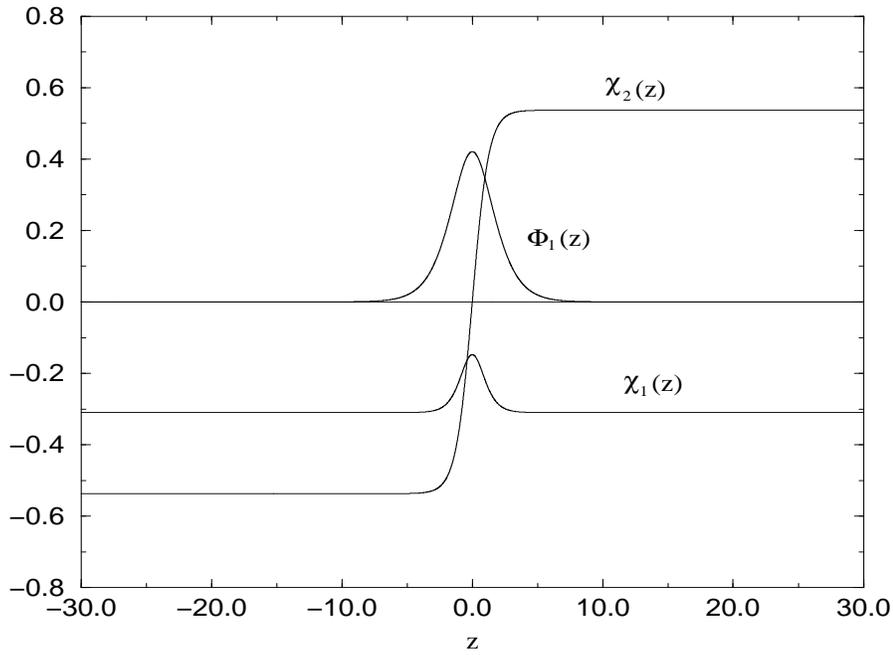,height=4in,width=5.5in}
\psfig{figure=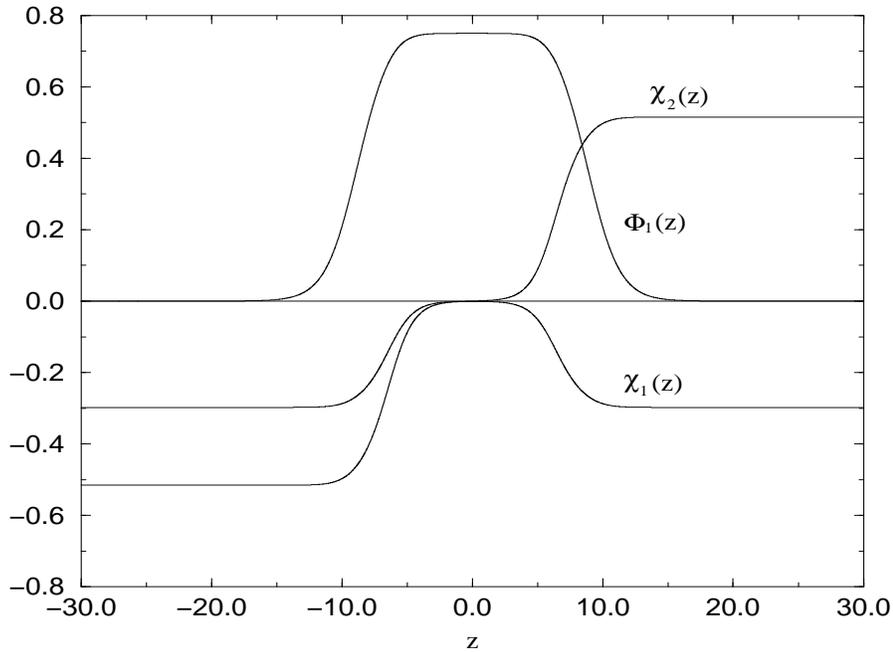,height=4in,width=5.5in}
\caption{\it Shape of a confined-confined domain wall. Deep in the confined 
phase (a) $\Phi_1(0) \neq 0$, i.e. the center of the wall has properties of the
deconfined phase. Close to the phase transition (b) the wall splits into two 
confined-deconfined interfaces with a complete wetting layer of deconfined 
phase between them.}
\end{figure}
Figure 1 shows a numerical solution of these equations for a domain wall 
separating two confined phases of type $(5)$ and $(6)$, i.e. with boundary
conditions $\Phi(\infty) = \Phi^{(5)}$, $\chi(\infty) = \chi^{(5)}$ and
$\Phi(-\infty) = \Phi^{(6)}$, $\chi(-\infty) = \chi^{(6)}$. Figure 1a 
corresponds to a temperature deep in the confined phase. Still, at the domain
wall the Polyakov loop is non-zero, i.e. the center of the domain wall shows
characteristic features of the deconfined phase. Figure 1b corresponds to
a temperature very close to the phase transition. Then the confined-confined
domain wall splits into two confined-deconfined interfaces and the deconfined
phase forms a complete wetting layer between them. The solutions of figure 1
have deconfined phase of type $(1)$ at their centers. Due to the $\Z(3)_c$
symmetry, there are related solutions with deconfined phase of types $(2)$ and 
$(3)$.

For the special values $d = a = 0$, $e = b$, $f = c$, $g = 2 c$ one can find an
analytic solution for a confined-deconfined interface. Combining two of these
solutions to a confined-confined interface, one obtains
\begin{eqnarray}
\Phi_1(z)&=&- \frac{1}{2} \Phi_0 [\tanh\alpha(z-z_0) - \tanh\alpha(z+z_0)], \ 
\Phi_2(z) = 0, \nonumber \\
\chi_1(z)&=&- \frac{1}{4} \chi_0 [2 + \tanh\alpha(z-z_0) - \tanh\alpha(z+z_0)],
\nonumber \\
\chi_2(z)&=&\frac{\sqrt{3}}{4} \chi_0 [\tanh\alpha(z-z_0) + 
\tanh\alpha(z+z_0)],
\end{eqnarray}
where $\Phi_0 = \chi_0 = - 3b/4c$ and $\alpha = - 3b/4\sqrt{c}$ (with
$b<0$). When $d = a = 0$, the critical
temperature corresponds to $e^4/f^3 = b^4/c^3$. Near criticality, where
$\Delta = e^4/f^3 - b^4/c^3$ is small, the above solution is valid up to order
$\Delta^{1/2}$, while now $e = b$ and $f = c$ are satisfied to order $\Delta$.
The width of the deconfined complete wetting layer,
\begin{equation}
2 z_0 = - \frac{1}{2 \alpha} \log \Delta + C,
\end{equation}
where $C$ is a constant, grows logarithmically as we approach the phase 
transition temperature. This is the expected critical behavior for interfaces 
with short-range interactions \cite{Tra92}. It would be interesting if complete
wetting could be studied in the framework of M-theory. If so, it should 
correspond to the unbinding of a pair of D-branes, which are tightly bound in 
the low-temperature phase.

Now we wish to explain why the appearance of the deconfined phase at the center
of the domain wall allows a QCD string to end there. We recall that an
expectation value $\langle \Phi \rangle \neq 0$ implies that the free energy 
of a static quark is finite. Indeed, the solution of figure 1 containing 
deconfined phase of type (1) has $\Phi_1(0) \neq 0$, such that a static quark 
located at the center of the wall has finite free energy. As one moves away
from the wall, the Polyakov loop decreases as
\begin{equation}
\Phi_1(z) \propto \exp(- F(z)/T) \propto \exp(- \sqrt{2a + 2 g \chi_0^2} \ z).
\end{equation}
Consequently, as the static quark is displaced from the wall, its free energy
$F(z)$ increases linearly with the distance $z$ from the center, i.e. the quark
is confined to the wall. The string emanating from the static quark ends on the
wall and has a tension
\begin{equation}
\sigma = \lim_{z \rightarrow \infty} \frac{F(z)}{z} = 
\sqrt{2a + 2 g \chi_0^2} \ T.
\end{equation}
Still, there are the other domain wall solutions (related to the one from above
by $\Z(3)_c$ transformations), which contain deconfined phase of types (2) and 
(3). One could argue that, after path integration over all domain wall 
configurations, one obtains $\langle \Phi \rangle = 0$. However, this is not 
true. In fact, the wetting layer at the center of a confined-confined domain 
wall is described by a two-dimensional field theory with a spontaneously broken
$\Z(3)_c$ symmetry. Consequently, deconfined phase of one definite type 
spontaneously appears at the domain wall. This argument does not apply to an
interface of finite area, e.g. to a bubble wall enclosing a finite volume of
confined phase of type (5) in a Universe filled with confined phase of type 
(4). Indeed, due to quantum tunneling the deconfined wetting layer at the 
surface of such a bubble would change from type (1) to types (2) and (3). As a
consequence, $\langle \Phi \rangle = 0$, such that a QCD string cannot end at 
the bubble wall. This observation is consistent with the $\Z(3)_c$ Gauss law 
applied to the compact surface of the bubble. Color flux entering the bubble 
through a string must exit it somewhere else and then go to infinity. 
Therefore, the static quark at the origin of the string has infinite free 
energy and is confined in the usual sense. The deconfined wetting layer of an 
infinite domain wall, on the other hand, can transport flux to infinity at a 
finite free energy cost. Thus, a static quark at a finite distance from the 
wall has a finite free energy and the string emanating from it can end on 
the wall. The above arguments imply that a QCD string can also end at a 
confined-deconfined interface at the high-temperature phase transition. This is
possible even in non-supersymmetric Yang-Mills theory.

We now understand that QCD strings can end on a confined-confined domain wall 
provided it is completely wet by deconfined phase. So far we have assumed that,
above the phase transition, confinement is lost and simultaneously chiral 
symmetry is restored. In that case, indeed there is complete wetting. Now let 
us consider two alternative scenarios for the phase transition. First, we 
assume that chiral symmetry is still restored at temperatures above the phase 
transition, but that the theory remains confining. In that case, complete 
wetting still occurs, but now the wetting layer consists of chirally symmetric 
confined phase, which has $\Phi = 0$. Under these conditions, a QCD string 
could not end on the wall. Next, we assume that, above the phase transition, 
the theory is deconfined, but that chiral symmetry remains broken, such that 
there are nine high-temperature phases. Assuming that the gluino condensate 
does not vary drastically across the transition, one can show that complete 
wetting then does not occur. Hence, again $\Phi = 0$ at the center of the 
domain wall and a QCD string could not end there. It should be noted that a 
confined-confined domain wall could still be incompletely wet, i.e. droplets of
deconfined phase could appear at the wall. Even in that case, a QCD string 
cannot end at a droplet. In contrast to a complete wetting layer, a droplet has
a finite volume. Hence, due to quantum tunneling a deconfined droplet of type 
(1) will turn into one of types (2) and (3). As a result, the average Polyakov 
loop vanishes for these configurations, and a static quark has infinite free 
energy even close to the wall. Given that QCD strings can end on the 
walls, as explained by Witten, we conclude that the two alternative scenarios 
for the phase transition can be ruled out. Interestingly, Witten's M-theory 
arguments in combination with our results for complete wetting suggest that 
$N=1$ supersymmetric QCD has a high-temperature phase transition in which 
simultaneously confinement is lost and chiral symmetry is restored.

In conclusion, we have investigated $\Z(3)_c \otimes \Z(3)_\chi$ symmetric
effective theories for the Polyakov loop and the gluino condensate. Due to
complete wetting, deconfined phase appears at a confined-confined domain wall.
Thus, close to the wall a static quark has a finite free energy and its 
string can end there. This is possible only when the wall is infinitely 
extended, because only then it can transport the quark's color flux to 
infinity. Hence, without reference to M-theory, we can understand why a QCD 
string can end on the wall.

\section*{Acknowledgements}

We are indebted to E. Farhi, J. Goldstone, K. Intrilligator, L. Randall, 
A. Shapere, M. Shifman, M. Strassler and E. Witten for very interesting 
discussions. The work of A. C. has been supported by the Direcci\'o
General de Recerca under a cooperative agreement between MIT and 
Generalitat de Catalunya. U.-J. W. wishes to thank the A. P. Sloan 
foundation for its support.

\end{document}